# Floppy hydrated salt microcrystals

Rozeline Wijnhorst[1], Menno Demmenie[1], Etienne Jambon-Puillet[1]†, Freek Ariese[2], Daniel Bonn[1], Noushine Shahidzadeh[1]*

[1]Institute of Physics, Van der Waals-Zeeman Institute, University of Amsterdam, Science Park 904, 1098 XH Amsterdam, The Netherlands.

[2]LaserLaB, Biophotonics and Medical Imaging, Vrije Universiteit Amsterdam, De Boelelaan 1081, 1081 HV Amsterdam, The Netherlands.

†Current address: Department of Chemical and Biological Engineering, Princeton University, NJ 08544 Princeton, USA.

*Corresponding author. Email: n.shahidzadeh@uva.nl.

**Summary**

**The crystalline structure of minerals due to the highly ordered assembly of its constituent atoms, ions or molecules confers a considerable hardness and brittleness to the materials[1,2]. As a result, they are generally subject to fracture. Here we report that microcrystals of natural inorganic salt hydrates such as sodium sulfate decahydrate ($Na_2SO_4 \cdot 10H_2O$) and magnesium sulfate hexahydrate ($MgSO_4 \cdot 6H_2O$) can behave remarkably differently: instead of having a defined faceted geometrical shape and being hard or brittle, they lose their facets and become soft and deformable when in contact with their saturated salt solution at their deliquescence point. As a result, the hydrated crystals simultaneously behaves as crystalline and liquid-like. We show that the observed elasticity is a consequence of a trade-off between the excess capillary energy introduced by the deformation from the equilibrium shape and viscous flow over the surface of the microcrystals. This surprising, unusual mechanical properties reported here reveals the role of the water molecules present in the crystalline structure. Although many compounds can incorporate water molecules in their crystalline frameworks, the relationship between the different hydrates and anhydrate crystal forms are still poorly investigated. Our results on the floppy behaviour of such crystalline structure reveals some unexplored properties of water molecules entrapped in the crystalline structure and can open novel routes for their application in various fields such as pharmaceutical sciences[3] , thermal energy storage[4–6] and even the traceability of water on Mars[7–9]**

## Main text

Minerals are natural materials that make up a large part of the earth and other planets[7–9] . Their chemical formula giving the types and proportions of the chemical elements and their crystal lattice defining the arrangement of atoms and the nature of bonding determine generally their physical properties. These physical properties and some other exceptional discovered properties of minerals such as fluorescence, magnetism or radioactivity, etc are of great technological significance for scientists and engineers for the design of novel materials.

For a typical crystalline solid, such as a salt, the spacing is of the order of a few ångström and the interaction energy several $k_BT$, which immediately leads to the conclusion that its elastic modulus is many GPa[1]. The consequence of this large modulus is that, when large stresses are applied to them, the crystal will break, preferentially along directions/planes corresponding to the lowest bond energy or along random directions [1,2]. It is well known that, due to the presence or absence of defects, the mechanical properties at small scales can differ from the macroscopic ones. However these effects are usually only visible at the nanoscale [10].

Here we show that microcrystals of natural inorganic salt hydrates such as sodium sulfate decahydrate ($Na_2SO_4 \bullet 10H_2O$) and magnesium sulfate hexahydrate ($MgSO_4 \bullet 6H_2O$) can behave remarkably differently instead of being hard, brittle and faceted, they become soft and deformable and lose their facets at their deliquescence point. In addition, we report 'self-healing' properties as indentations on the crystal surface spontaneously and rapidly close themselves under the action of surface tension. This unusual behavior appears to be related to the number of water molecules present in the crystalline structure and their mobility; in some hydrated salts the water is stoichiometrically bound into a crystal but not directly bonded to the central metal cation.

We study the formation of sodium sulfate decahydrate ($Na_2SO_4 \cdot 10H_2O$) and magnesium sulfate hexahydrate ($MgSO_4 \cdot 6H_2O$) by the deliquescence of the anhydrous form of these salts using electron and optical microscopy as well as confocal Raman micro spectroscopy. Deliquescence is the absorption of water vapor by the crystals from the atmosphere until it dissolves and forms a solution. First, the anhydrous form of sodium sulfate is obtained by drying millimetric mirabilite crystals precipitated in salt solution at room temperature. Mirabilite or the sodium sulfate decahydrate also referred to as Glauber's salt [11], is the thermodynamically stable state in contact with a saturated sodium sulfate solution at room temperature (**Extended Data Fig. 1**) [12]. However, once the crystals are taken out of solution, the water in the crystalline structure evaporates easily at room temperature, and subsequently the initially transparent hydrated crystal transforms into a white porous structure made up of micro crystallites, while retaining the overall shape of the original millimeter-sized mirabilite crystal (**Fig. 1a,b**). Scanning Electron microscopy imaging shows that this dehydrated white porous structure is made of an assembly of microcrystals of thenardite (**Fig. 1d**), as identified by confocal Raman microscopy with an $SO_4^{2-}$symmetric stretch peak at 994 $cm^{-1}$ (**Extended Data Fig. 2**) in agreement with previous work [13,14]. It is this porous structure that is subsequently rehydrated by deliquescence.

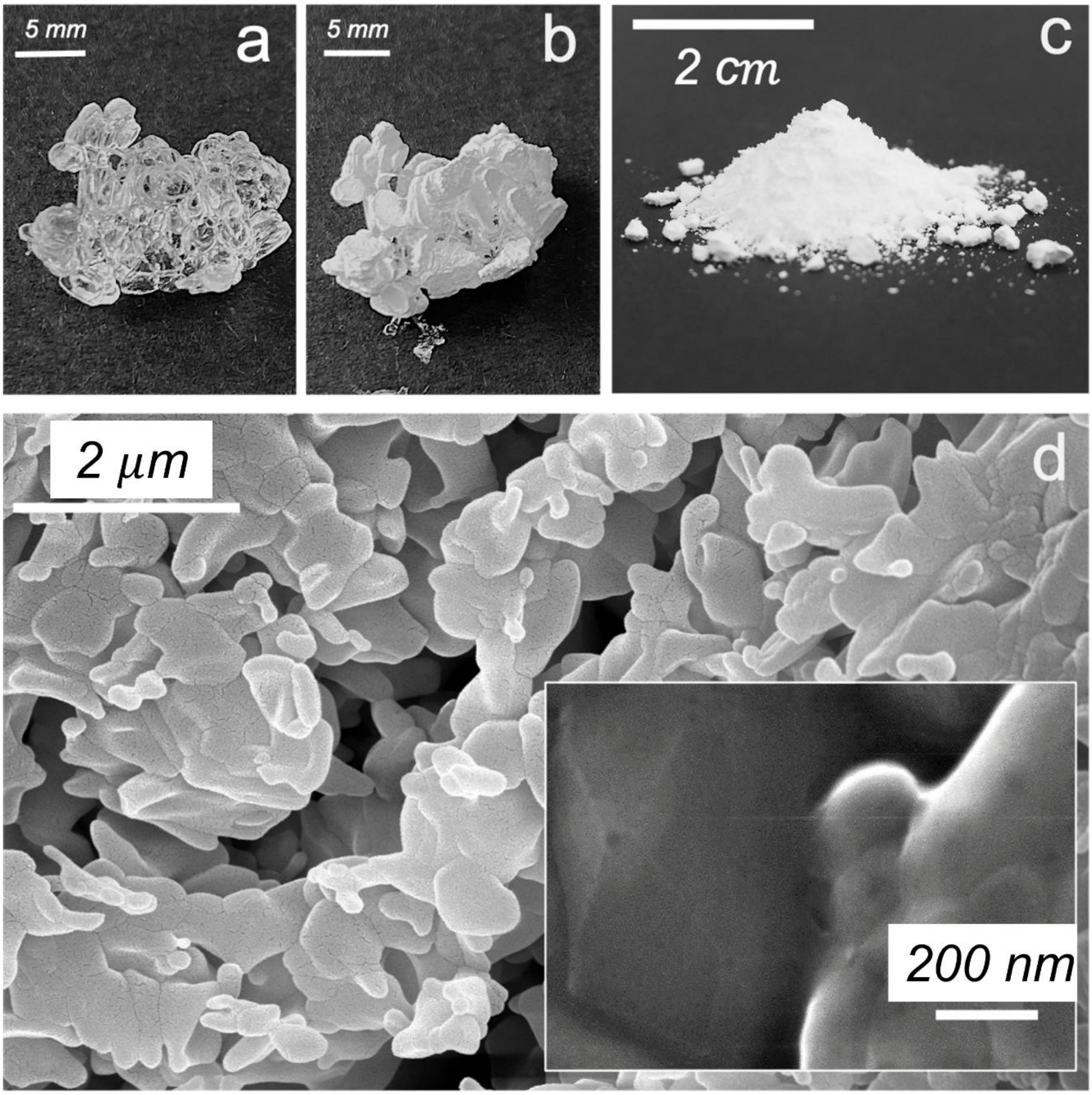

**Fig. 1. Mirabilite ($Na_2SO_4.10H_2O$) and thenardite ($Na_2SO_4$) crystals**. (**a**) typical mirabilite crystals precipitated out of a sodium sulfate solution, and (**b**) mirabilite crystals dried (dehydrated) at T=21ºC (**c**) Powder of the dried crystals (**d**) Electron microscopy images of the dried crystal consisting of an assembly of nanocrystals of thenardite.

To do so, a small amount of the anhydrous powder is placed under a polarizing light microscope in a climate chamber at a relative humidity above the equilibrium relative humidity $RH_{eq} \sim 96\%$ of this salt (**Extended Data Fig. 1**) [4]. The progressive moisture sorption and dissolution of the thenardite nanocrystals results in the gradual formation of a salt solution in which subsequently the growth of new crystals of mirabilite can be observed (**Fig. 2a,b**). This is confirmed by analysis of sequential Raman spectra of the process, that reveals an overlay of the thenardite (994 $cm^{-1}$) and mirabilite (989 $cm^{-1}$) spectra; the intensity of the anhydrous polymorph (thenardite) decreases in time due to the deliquescence, whereas that of the mirabilite spectrum increases with the formation and growth of more mirabilite microcrystals (**Fig. 2c**) ; the latter can also entrap undissolved thenardite as impurities during their growth (seen as black dots in **Fig. 2**) in their structures during the growth. The process thus leads to have large amount of microscale size mirabilite crystals instead of a few large millimetric ones as seen **in Fig1a**.

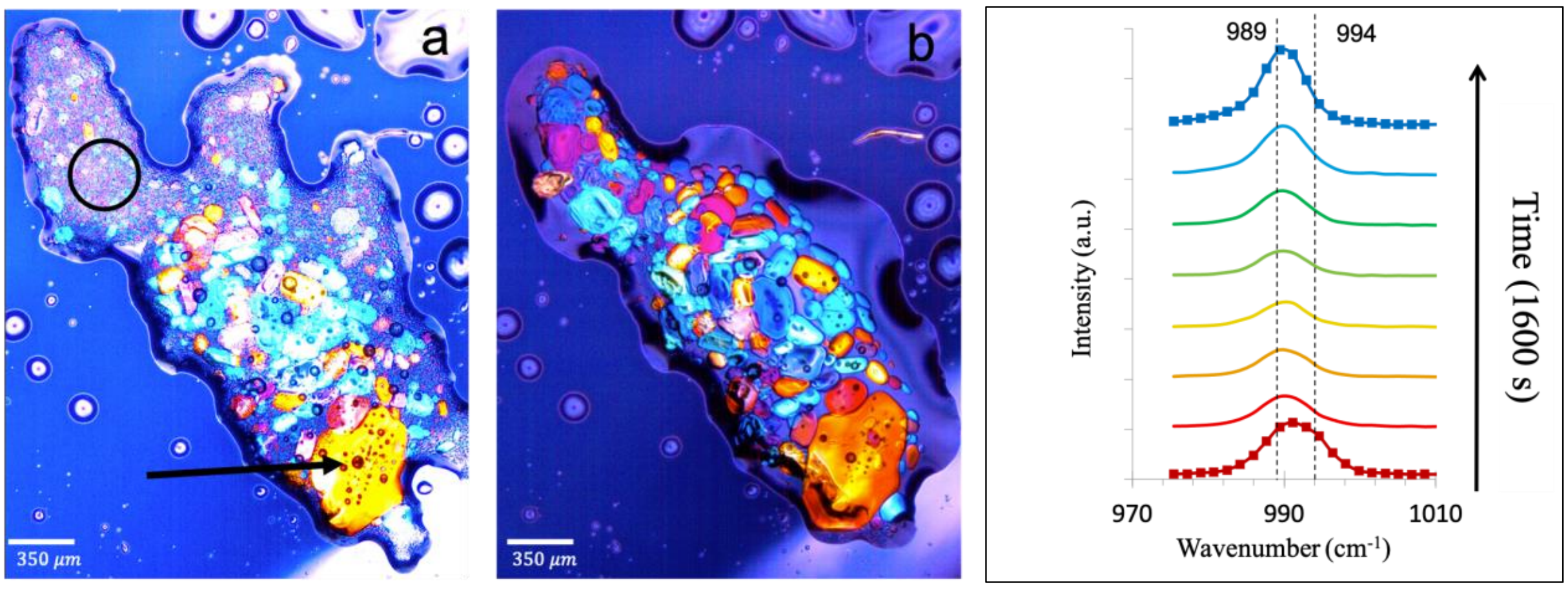

**Fig. 2. Gradual deliquescence of thenardite ($Na_2SO_4$) polycrystals and rapid precipitation of mirabilite ($Na_2SO_4{\cdot}10H_2O$) micro crystals. (a-b)** Microscopy images made with cross polarizers of the deliquescence process of the dehydrated powder. The colors that are seen with cross polarized lights highlights the fact that what we are seeing is not a liquid structure. Initially, both polymorphs (mirabilite and thenardite) coexist as shown by the Raman spectra of the region circled in black **(b)** After 660 s, most of the thenardite nanocrystals are dissolved; further water vapour absorption will cause the mirabilite crystals to deliquesce in turn. The arrow shows entrapped remaining thenardite nanocrystals (black dots) in the mirabilite crystals. **(c)** Raman spectra of the evolution of the $SO_4$ symmetric stretch peak during deliquescence of thenardite powder; the dotted lines indicate the peak maxima of thenardite (994 $cm^{-1}$) and mirabilite (989 $cm^{-1}$) (see **Extended Data Fig. 2**).

Generally, the higher the number of water molecules in the crystalline structure of a given inorganic hydrated crystal, the lower its solubility at a given temperature(phase diagrams from the literature are shown in **Extended Data Fig. 1).** 10,13,14,15 In our experiment, the salt solution formed by deliquescence of thenardite nanocrystals is therefore highly concentrated with respect to the formation of the decahydrate form (**Extended Data Fig. 1**); the resulting supersaturation ($S \sim m_{\text{thenardite}} / m_{\text{mirabilite}} \sim 2.5$ at 21ºC) induces the rapid precipitation of the large amount of mirabilite microcrystals. Their spectacular-colored appearance in polarizing light microscopy (**Fig. 2)** and Raman identification of these crystals as mirabilite (**Fig. 3b**) rules out the formation of an intermediate metastable amorphous phase or dense liquid droplets on the time scale of our experiments. The mirabilite microcrystals are observed to grow only until a certain size, until the surrounding salt solution concentration drops to the saturation concentration for mirabilite (**Extended Data Fig. 1**). Beyond this threshold, the precipitated mirabilite crystals will then dissolve in their turn, as the deliquescence process (i.e., the water vapour uptake) continues.

The remarkable observation is that, although mirabilite has a clear monoclinic crystal structure, the precipitated small mirabilite crystals close to their deliquescence point exhibit no facets at all and appear very rounded. In addition, they are observed to be soft; this can be most clearly observed in regions with a high local density of crystals: the microcrystals are observed to deform each other and adapt to the shape of the other neighboring crystals, much like liquid droplets in an emulsion (**Fig. 3a**). Once the strain imposed by the presence of other crystals is removed, the deformed crystals regain their non-faceted spherical shape (**Fig. 3c-e**).

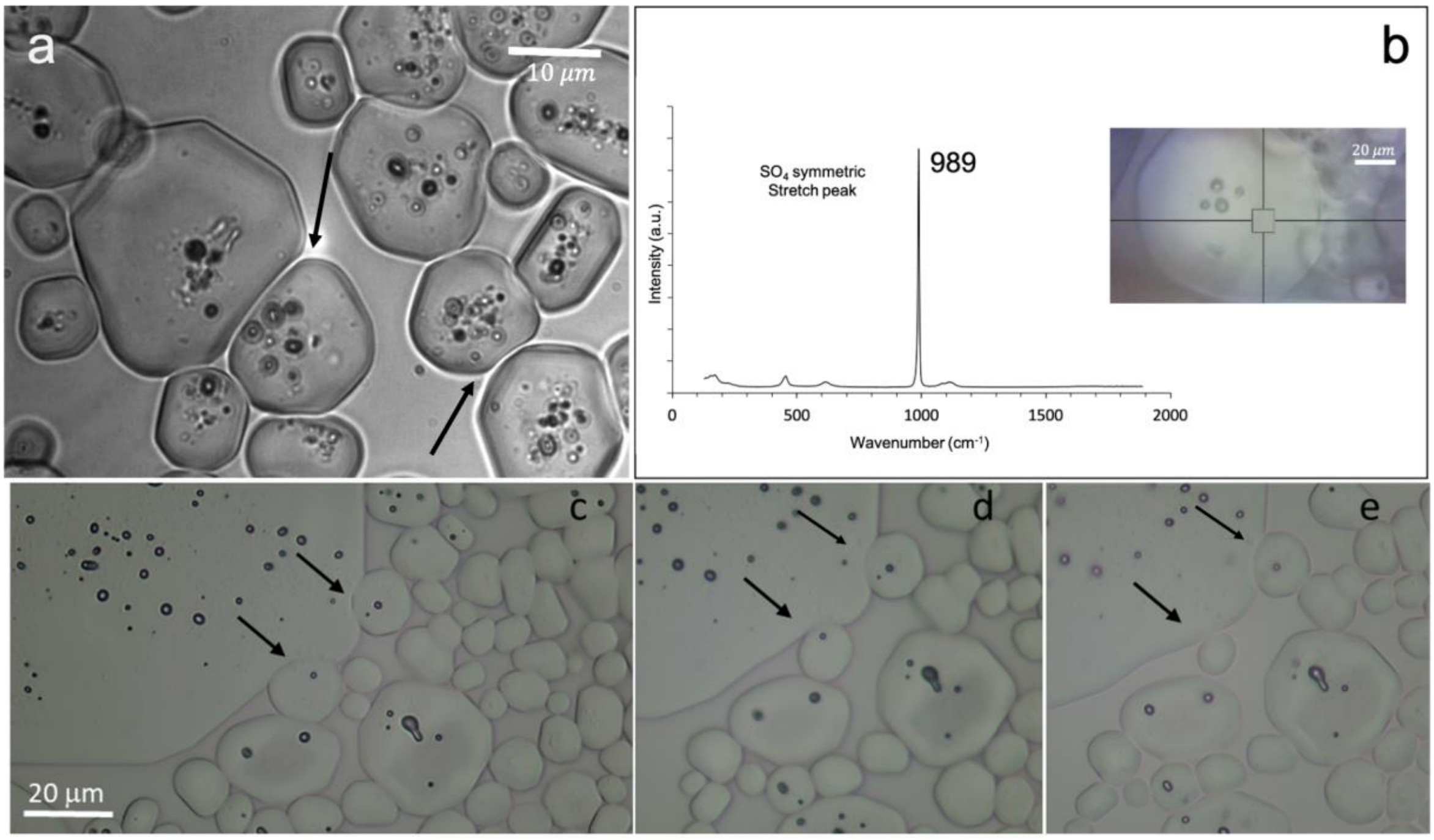

**Fig. 3. Floppiness of mirabilite micro crystals surrounded by sodium sulfate solution at their deliquescence point**. (**a**) At the deliquescence point, non-faceted mirabilite micro crystals surrounded by sodium sulfate solution are deformable and can adapt their shape with respect to the neighboring crystals. (**b**) Raman spectrum of the crystals identified as mirabilite. (**c**) the arrows show the local deformation of a large mirabilite crystal by two neighboring small crystals like pushing on a soft ball; **(d-e)** The crystals show an elastic response once the strain is removed indicated with the black arrows.

One way to interpret these results would be to assume that the mobility of the molecules at the crystalline surface is very high. In agreement with this hypothesis, we observe in a separate set of experiments that any indentations at the crystal surface can heal spontaneously. Such indentations appear during the deliquescence of the mirabilite microcrystals when entrapped thenardite nanocrystals (seen as black dots in **Fig. 2 and 3**) are released into the solution and quickly dissolve in the surrounding salt solution, which is not fully saturated with respect to the thenardite polymorph (**Fig. 4a,b**). We observe that the holes thus formed close up spontaneously.

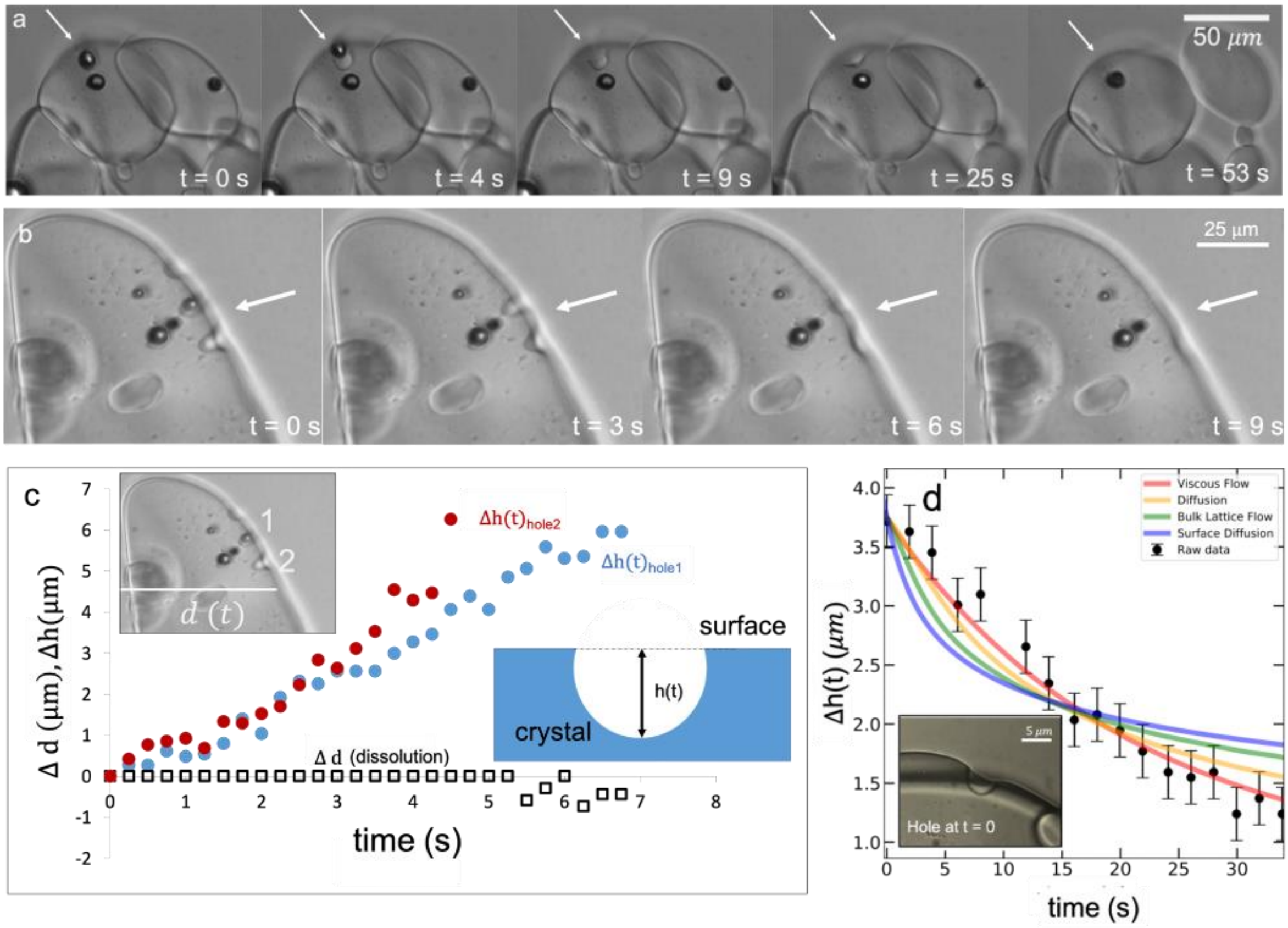

**Fig. 4. Self-healing behavior at the surface of mirabilite crystals.** (**a-b**) evolution of surface defects on two different crystals, indicated by arrows. During the deliquescence, while the entrapped thenardite dissolves, the indentation at the surface of the hydrated crystal is seen to level out rapidly. (**c**) The temporal evolution of the levelling out $\Delta h(t)= h(t)- h_0$ of two holes (filled circles) and the dissolution $\Delta d(t)= d(t)- d_0$ of the same crystal (open squares) (**d**) Comparison between the measured depth of a hole $h(t)$ levelling towards a flat surface (symbols) and the Mullins model (solid lines) describing the four possible mechanisms explaining the levelling. The model describing a viscous flow (red curve) shows the best agreement with the data.

One can imagine that the levelling of the holes could simply be due to the dissolution of the mirabilite crystal because of the deliquescence. However, the comparison of the dissolution speed of the crystals $d(t)$ with the temporal evolution of the depth of several holes $h(t)$ of different sizes on various mirabilite crystals shows that the dissolution is much too slow to account for the disappearance of the holes (**Fig. 4c**). The self-healing behaviour of the crystal surface has thus another physical origin. To determine what mechanism is responsible for the hole closing, we use Mullins model[17–19] which solves the levelling of a hole due to four different possible mechanisms: (1) viscous-capillary flow of the crystal, (2) diffusion of the salt from the solution into the holes, (3) a bulk lattice rearrangement, and (4) a rearrangement of surface molecules by surface diffusion. We compare this model to the full experimental hole profiles measured during the levelling in (**Extended Data Fig. 3)** and to the hole depth (**Fig. 4d**) and find that the viscous-capillary flow mechanism is the only one able to describe the data (for details on the model, equation and the comparison with experiments, see Materials and Methods in the Extended Data). The constant in the equation represents a capillary velocity; its value is approximately $0.1\mu ms^{-1}$ for the hole shown in Figure 4d and does not vary much between different experiments. Therefore, the self-healing, i.e., the relaxation of the holes is driven by the gradients in Laplace pressure along the crystal-solution interface; the excess capillary energy introduced by the curvature of the hole with respect to the flat equilibrium state is dissipated by viscous flow into the hole that leads to the observed self-healing. The driving force for the healing is thus clearly the surface tension that requires the crystals to be spherical.

This unusual soft, non-faceted and self-healing behavior at the deliquescence point is not specific to sodium sulfate. Repeating the same experiments for micro crystals of magnesium sulfate hexahydrate ($MgSO_4 \cdot 6H_2O$), we observed a similar behavior under the polarizing and Raman confocal microscope (**Extended Data Fig. 4**). The process is, however, much faster since the deliquescence of magnesium sulfate starts at a lower equilibrium relative humidity $RH_{eq}$= 85% [16], which induces a faster water vapor sorption rate in comparison with sodium sulfate in our climate chamber (where the relative humidity of the chamber was fixed at RH~100%). In addition, the magnesium sulfate hexahydrate microcrystals were smaller in size, which prevented us from measuring the self-healing process quantitatively with our experimental setup as the indentations were too small and closed too rapidly. Nonetheless, the magnesium sulfate salt behaved qualitatively like the sodium salt described above.

Our observations of the self-healing and softness of hydrated crystals are reminiscent of the case of Helium-4 crystals close to zero temperature, which under the application of extremely small stresses present a giant plasticity due to the motion of crystalline defects. This leads, amongst other things, to the 'dripping' of such crystals reported to behave liquid-like rather than solid-like [20–22] due to the bulk lattice motion ('model 3' in our terminology)[20–22]
In our case of hydrated salts, the deformability of the microcrystals at the deliquescence point is mainly caused by capillary-driven viscous flow at the surface of the crystal. This governs the self-healing dynamics of the crystals and shows why the molecules in the crystalline lattice appear to be mobile, leading to spherical and deformable crystals on our experimental time scale. The time scale for the viscous flow limits the size range over which the floppiness can be observed. The comparison of the behavior between the sodium and magnesium salt is interesting; one hypothesis could be that the higher numbers of water molecules in the hydrated crystal, water can be more easily exchanged between the crystal and the layer adjacent to it. Second, since the magnesium is bivalent, the water is likely to be more strongly bound, again decreasing the exchange rate with the exterior water. The ease of exchange of the water molecule in the crystalline structure with the surrounding is also indicated by the fact that this water of hydration evaporates more readily from hydrated crystals with a relatively high number of water molecules [4,5,23]. For the sodium sulfate decahydrate this readily happens at room temperature, whereas for hydrated salts with fewer water molecules in the lattice the water evaporates at higher temperatures (for magnesium sulfate hexahydrate T>60$^o$C, for magnesium sulfate monohydrate and calcium sulfate dihydrate T>150$^o$C). The same reasoning then also explains why the softness is not observable for the anhydrous form of the salt.
Another important observation in this respect is that when the relative humidity is decreased below the equilibrium relative humidity $RH_{eq}$ (i.e., below the deliquescence point), the unfaceted hydrated crystals will lose their softness, facets appear again and increase in size over time. Thus, the surface mobility only appears at the deliquescence point (i.e., at a certain RH for a given salt); perhaps in analogy to the roughening transition that appears at a fixed temperature for certain crystals, and above which the facets also disappear.

The hydrated salts that we study here are not only abundant and ubiquitous on the surface of the Earth but are also present on Mars[8,9,15]. In fact, their discovery on the sediments of Mars provided vital information on the hydrogeologic history of other planets. In addition, they are an important class of natural materials that are currently investigated both as thermochemical materials and phase change materials (PCM) and for energy storage purposes because of their large potential storage capacity (~ 1-3 $GJ/m^3$) and the easily accessible temperature range that they cover. Our results reveal an unexplored fundamental property of inorganic hydrated salts, which can open new routes for optimizing their performance with improved control over their geometry and size. On the other hand, such materials have tremendous potential for applications where the key parameter is relative humidity rather than temperature. More generally, the conception of novel *soft microcrystalline materials* with peculiar macroscopic properties with respect to their shape, colour, and luminescence as ‘flexible responsive systems’ remains an underdeveloped area that deserves further exploration.[24–26]

## Main references

## Figure legends

**Fig. 1. Mirabilite ($Na_2SO_4.10H_2O$) and thenardite ($Na_2SO_4$) crystals**. (**a**) typical mirabilite crystals precipitated out of a sodium sulfate solution, and (**b**) mirabilite crystals dried (dehydrated) at T=21ºC (**c**) Powder of the dried crystals (**d**) Electron microscopy images of the dried crystal consisting of an assembly of nanocrystals of thenardite

**Fig. 2. Gradual deliquescence of thenardite ($Na_2SO_4$) polycrystals and rapid precipitation of mirabilite ($Na_2SO_4{\cdot}10H_2O$) micro crystals. (a-b)** Microscopy images made with cross polarizers of the deliquescence process of the dehydrated powder. The colors that are seen with cross polarized lights highlights the fact that what we are seeing is not a liquid structure. Initially, both polymorphs (mirabilite and thenardite) coexist as shown by the Raman spectra of the region circled in black **(b)** After 660 s, most of the thenardite nanocrystals are dissolved; further water vapour absorption will cause the mirabilite crystals to deliquesce in turn. The arrow shows entrapped remaining thenardite nanocrystals (black dots) in the mirabilite crystals. **(c)** Raman spectra of the evolution of the $SO_4$ symmetric stretch peak during deliquescence of thenardite powder; the dotted lines indicate the peak maxima of thenardite (994 $cm^{-1}$) and mirabilite (989 $cm^{-1}$) (see **Extended Data Fig. 2**).

**Fig. 3. Floppiness of mirabilite micro crystals surrounded by sodium sulfate solution at their deliquescence point**. (**a**) At the deliquescence point, non-faceted mirabilite micro crystals surrounded by sodium sulfate solution are deformable and can adapt their shape with respect to the neighboring crystals. (**b**) Raman spectrum of the crystals identified as mirabilite. (**c**) the arrows show the local deformation of a large mirabilite crystal by two neighboring small crystals like pushing on a soft ball; **(d-e)** The crystals show an elastic response once the strain is removed indicated with the black arrows.

**Fig. 4. Self-healing behavior at the surface of mirabilite crystals.** (**a-b**) evolution of surface defects on two different crystals, indicated by arrows. During the deliquescence, while the entrapped thenardite dissolves, the indentation at the surface of the hydrated crystal is seen to level out rapidly. (**c**) The temporal evolution of the levelling out $\Delta h(t)= h(t)- h_0$ of two holes (filled circles) and the dissolution $\Delta d(t)= d(t)- d_0$ of the same crystal (open squares) (**d**) Comparison between the measured depth of a hole $h(t)$ levelling towards a flat surface (symbols) and the Mullins model (solid lines) describing the four possible mechanisms

explaining the levelling. The model describing a viscous flow (red curve) shows the best agreement with the data.

## Methods

### Recrystallization of the anhydrous $Na_2SO_4$ nanocrystals in the form of powder

First, sodium sulfate solutions were prepared by dissolving anhydrous thenardite $Na_2SO_4$ (Sigma Aldrich; 99.8% purity) little by little while stirring in Millipore water ($\rho \approx 18.2$ MΩ.cm) till a concentration of 18.6 wt % or 1.6 molal; beyond this concentration the solution becomes turbid due to the precipitation of mirabilite crystals ($Na_2SO_4 \cdot 10\ H_2O$) from the supersaturated solution (concentration at saturation being $m_s$= 1.4 molal). The stirring is then stopped and the bottle is kept at rest for one day at room temperature. Some precipitated mirabilite crystals were subsequently removed from the saturated salt solution and dried at ambient temperature and relative humidity RH~45% of the lab. The stability of the hydrated/anhydrous phase in open air depends on the temperature and relative humidity. Upon drying and dehydration, the transparent mirabilite crystals become white and crumbly while keeping the shape of the initial hydrated crystals (see **Fig. 1ab**). The composition and the morphology of the white structure has been identified by Raman spectroscopy and high-resolution scanning electron microscopy as being an assembly of thenardite nanocrystals.

### Deliquescence of anhydrous $Na_2SO_4$ nanocrystals

The anhydrous $Na_2SO_4$ powder was placed on a clean glass slide and compressed to minimize spaces between the powder particles. The glass slide is then placed into a controlled mini climatic chamber under the microscope and subjected to an increasing relative humidity. The temporal physical changes of the powder due to moisture adsorption was followed under the Leica DM IRM inverted microscope equipped with a 800 x 600 pixels and 8-bit sensitivity CCD camera. The images of the ensuing process were taken at different time intervals at 10x, 20x, 40x, and 60x magnifications. The images were subsequently analyzed using the Fiji ImageJ software.

### Recrystallization of the anhydrous $MgSO_4$ nanocrystals in the form of powder

A similar procedure as for the recrystallization of the $Na_2SO_4$ was followed.
First, magnesium sulfate solutions were prepared by dissolving anhydrous magnesium sulfate (Sigma Aldrich; 99.5% purity) little by little while stirring in Millipore water ($\rho \approx 18.2$ MΩ.cm) to obtain a supersaturated solution (concentration at saturation being $m_s$= 3.8 molal). The stirring was then stopped and the bottle was kept at rest for one day at room temperature. Droplets of the solution were put on a Corning glass slide (borosilicate) and were dried at T = 60°C. After ~24 hours the dried droplets were removed from the oven and crushed to get a white powder which was stored at RH = 20% to ensure the $MgSO_4$ stayed anhydrous. The composition and the morphology of the powder has been identified by Raman spectroscopy and high-resolution scanning electron microscopy as being anhydrous $MgSO_4$ crystals.

### Deliquescence of anhydrous $MgSO_4$ nanocrystals

The anhydrous $MgSO_4$ powder was placed on a clean glass slide. The glass slide was then placed into a controlled mini climatic chamber under the microscope and subjected to an increasing relative humidity. The temporal physical changes of the powder due to moisture adsorption were followed under the Leica DM IRM inverted microscope equipped with a 800 x 600 pixels and 8-bit sensitivity CCD camera. The images of the ensuing process were taken at different time intervals at 10x, 20x, 40x, and 60x magnifications. The images were subsequently analyzed using ImageJ software. The floppy crystals appeared at RH ~ 85%.

### Confocal Raman spectroscopy

For chemical identification of elements and various polymorphs of $Na_2SO_4$ and $MgSO_4$ during the deliquescence, we applied Raman microscopy using a Renishaw InVia spectrometer with a 532 nm laser calibrated with a silicon mono crystal. First, we took reference spectra in the fingerprint area of each compound: the mirabilite $Na_2SO_4 \bullet 10H_2O$, the anhydrous $Na_2SO_4$ powder obtained after dehydration and 1.4 molal saturated solution, using 20x and 50x magnification objectives, backscattering geometry and 100% laser power. These spectra help to distinguish different phases as the wavenumber position and number of peaks differ for each phase. The peak values of the reference spectra were compared to the literature values and an excellent agreement was found[12].
We next took sequential Raman spectra to characterize the full deliquescence process and the transformation of different phases. The deliquescence was carried out in a similar mini climatic chamber as used with the inverted Leica microscope, but with an aluminum slide instead of a glass slide at the bottom. Furthermore, we also did single measurements on the hydrated crystals that appeared during deliquescence.

**Statistical test to determine the best fitting model for the hole-levelling out**

In order to determine the driving mechanism behind the levelling out of the holes, we compared the data of the measurements with four candidate models: (1) relaxation by a viscous flow, (2) a diffusion process, (3) a bulk lattice flow, and (4) a rearrangement of surface molecules by surface diffusion. The theoretical framework for these mechanisms is given by Mullins.[17,18] He showed that for a one-dimensional sinusoidal indentation $u_0 \sin(kx)$, the profile evolves as $U(x,t) = u(t)\sin(kx)$, with *x* the direction perpendicular to the hole, and *t* the time and $u(t)$ the amplitude which is described by:

$$\frac{\partial u}{\partial t} = -C_n(T)k^n u,$$

Where $C_n(T)$ is a temperature-dependent prefactor, $k$ the wavenumber of the profile and *n* an integer that represents the levelling physical mechanism:

- $n = 1$: a viscous-capillary flow, for an incompressible fluid and a low Reynolds number.
- $n = 2$: a dissolution-precipitation process, driven by the Kelvin equation that states that a curved surface has a higher saturation concentration and therefore can precipitate ions dissolved in its vicinity.
- $n = 3$: a bulk diffusive lattice flow, can be caused by interstitial or substitutional mechanisms in the bulk of the salt crystal.
- $n = 4$: surface diffusion, a rearrangement of surface molecules driven by the gradient of the curvature of the profile.

Since Mullins also showed that the sum of two solutions is again a solution, we decomposed the initial experimental profile at $t = 0$ as a Fourier sum and propagate each component individually in time (where we used a cut-off after the 90 lowest-spatial-frequency components). We found that the experimental data showed the best agreement with the first order model describing the viscous flow process. This was obtained by a statistical test using chi-square minimization between the u-values of the data and the four candidate mechanisms. Note that in the simplification of the model the small slope approximation has been applied ($du/dx \ll 1$) for both the projection of the molecular velocity and the curvature of the hole.

**Methods references**

**Acknowledgements**
Thanks are due to Bram Mooij and Pien Bouman (LaserLaB) and Paul Kolpakov (UvA) for help with the Raman measurements.

**Author contributions**

Conceptualization: RW, NS

Methodology: RW, MD, DB, EJP, FA, NS

Investigation: RW, MD, DB, EJP, FA, NS

Visualization: RW, MD, EJP, NS

Project administration: NS

Supervision: NS, DB, FA

Writing – original draft: RW, MD, DB, NS

Writing – review & editing: RW, MD, DB, EJP, FA, NS

**Competing interest declaration**
Authors declare that they have no competing interests

**Additional information (containing supplementary information line, corresponding author line)**

**Supplementary Information** is available for this paper.
Correspondence and requests for materials should be addressed to Prof. dr. Noushine Shahidzadeh.

**Extended data figure/table legends**

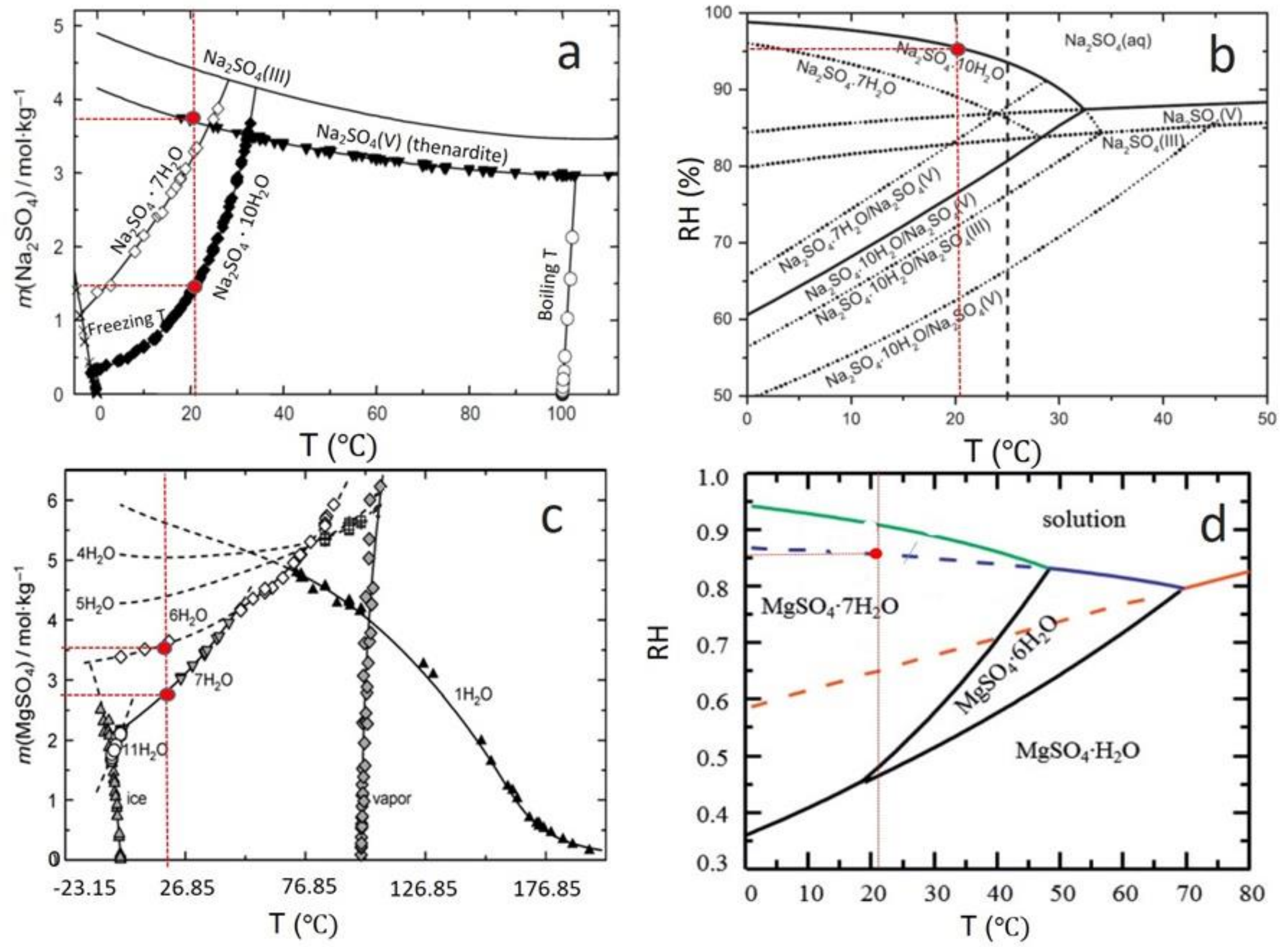

**Extended Data Figure 1 | Phase diagrams of sodium sulfate and magnesium sulfate from the literature**. Solubility phase diagram *(10)* (**a**) and humidity phase diagram *(13)* (**b**) of sodium sulfate Mirabilite ($Na_2SO_4$•10 $H_2O$), hepta hydrate (($Na_2SO_4$ • 7 $H_2O$), thenardite phase III and V ($Na_2SO_4$). Solubility phase diagram (*14*) (**c**) and humidity phase diagram (*15*) (**d**) of magnesium sulfate Epsomite ($MgSO_4$•7$H_2O$), hexahydrate ($MgSO_4$•6$H_2O$), kieserite ($MgSO_4$•$H_2O$). The red dotted lines represent our experimental conditions.

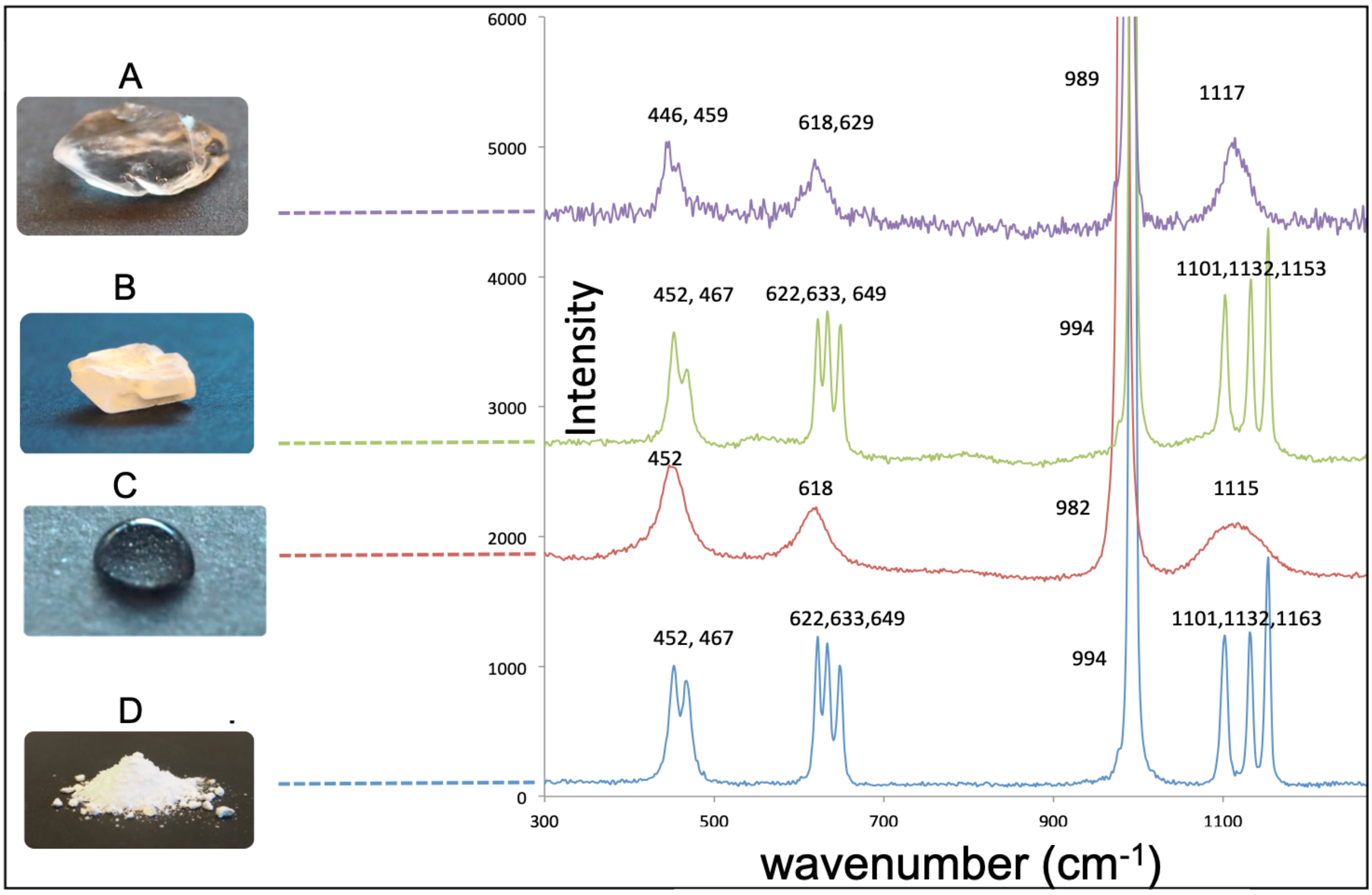

**Extended Data Figure 2 | Raman spectra of different phases of sodium sulfate.** (**A**) Mirabilite crystal (decahydrate of $Na_2SO_4$) (**B**) Dried $Na_2SO_4$ crystal, identified as thenardite (anhydrous $Na_2SO_4$). (**C**) Aqueous $Na_2SO_4$ solution. (D) Crushed dried $Na_2SO_4$ crystal, identified as thenardite (anhydrous $Na_2SO_4$). Spectra are vertically offset for clarity; intensities in arbitrary units.

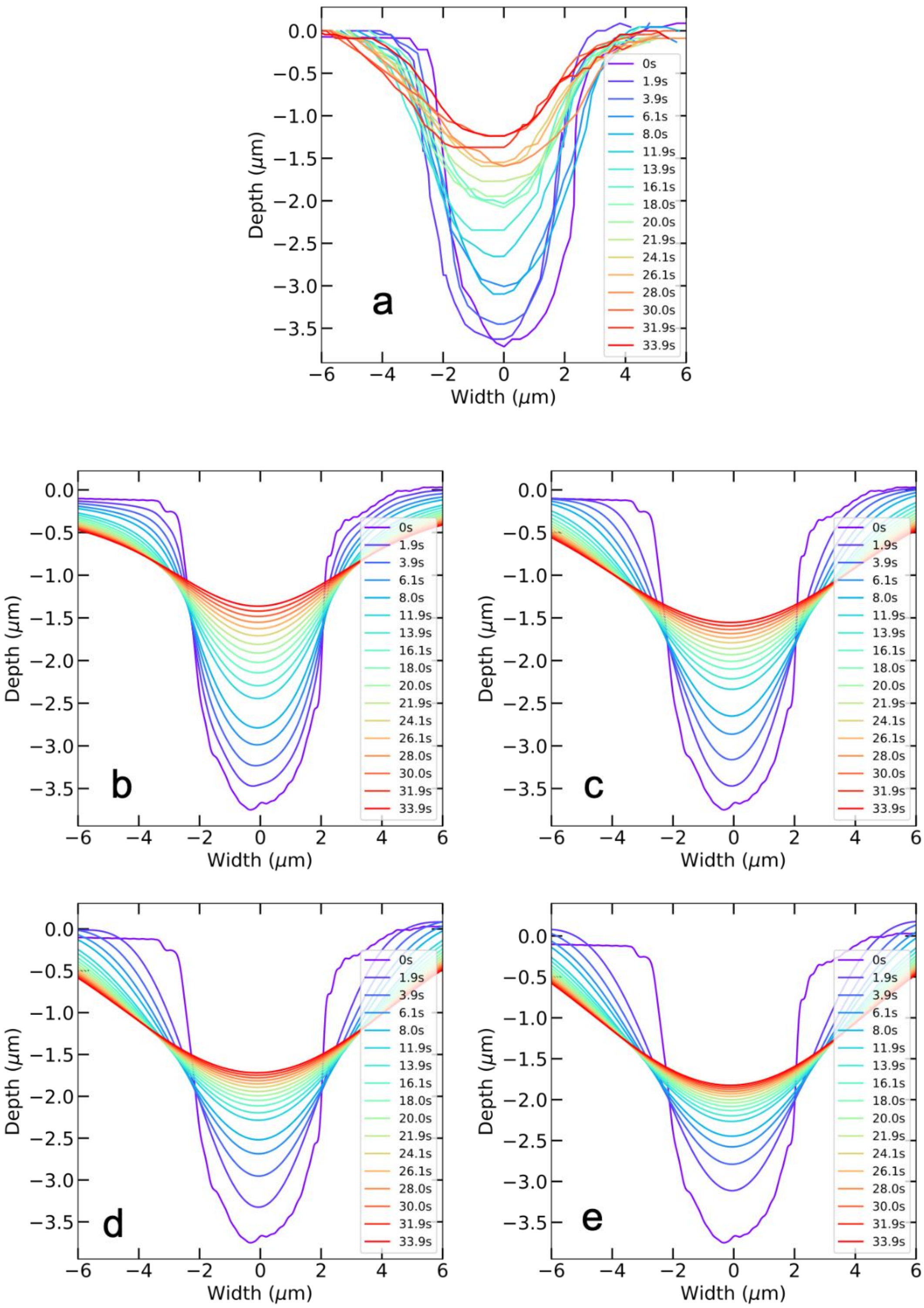

**Extended Data Figure 3 | Comparison of the four models to the mirabilite self-healing data. (a)** Raw data; **(b)** Effective flow model (*n* = 1); **(c)** Diffusion model (*n* = 2); **(d)** Bulk lattice flow model (*n* = 3); **(e)** Rearrangement of surface molecules by surface diffusion model (*n* = 4).

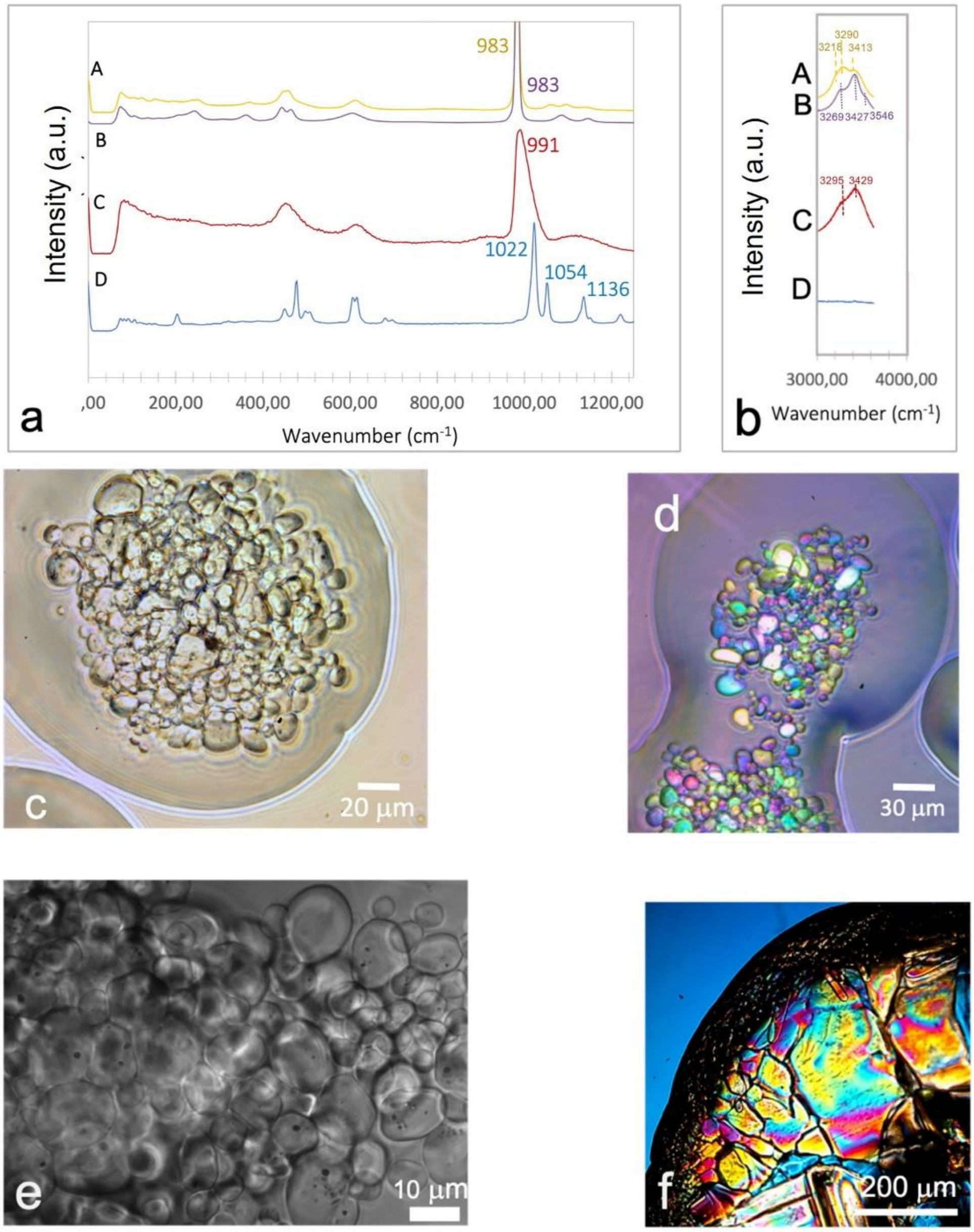

**Extended Data Figure 4 | Raman and optical microscopy of magnesium sulfate. (a,b)** Raman spectrum of the 'floppy' magnesium sulfate crystal (A) compared to that of the hexahydrate crystal (B), a saturated solution of magnesium sulfate (C), and anhydrous magnesium sulfate powder (D). **(c-e)** Floppy magnesium sulfate crystals under optical and light polarizing microscope. **(f)** Hexahydrate of magnesium sulfate in a dried droplet under the light polarizing microscope.

**Extended Data Table 1 | Levelling velocity of holes in mirabilite crystal walls for holes of different depths**

| Hole depth (μm) | Levelling velocity (μm/s) |
|---|---|
| 4.2 | 0.205 ± 0.03 |
| 1.8 | 0.096 ± 0.03 |
| 4.2 | 0.535 ± 0.08 |
| 12.1 | 0.143 ± 0.01 |

This value is determined by the best fitting value $C_1(T)$ of the first order model.